# A new seeding technique for the reliable fabrication of large, SmBCO single grains containing silver using top seeded melt growth

Y-H Shi, A R Dennis and D A Cardwell

Department of Engineering, University of Cambridge, Trumpington Street, Cambridge, CB2 1PZ, UK

Email: ys206@cam.ac.uk

**Abstract**

Silver (Ag) is an established additive for improving the mechanical properties of single grain, (RE)BCO bulk superconductors (where RE = Sm, Gd and Y). The presence of Ag in the (RE)BCO bulk composition, however, typically reduces the melting temperature of the single crystal seed in the top seeded melt growth (TSMG) process, which complicates significantly the controlled nucleation and subsequent epitaxial growth of a single grain, which is essential for high field engineering applications. The reduced reliability of the seeding process in the presence of Ag is particularly acute for the SmBCO system, since the melting temperature of SmBCO is very close to that of the generic NdBCO(MgO) seed. SmBCO has the highest superconducting transition temperature, $T_c$, and exhibits the most pronounced "peak" effect at higher magnetic field of all materials in the family of (RE)BCO bulk superconductors and, therefore, has the greatest potential for use in practical applications (compared to GdBCO and YBCO, in particular). Development of an effective seeding process, therefore, is one of the major challenges of the TSMG process for the growth of large, high quantity single grain superconductors. In this paper, we report a novel technique that involves introducing a buffer layer between the seed crystal and the precursor pellet, primarily to inhibit the diffusion of Ag from the green body to the seed during melt processing in order to prevent the melting of the seed. The success rate of the seeding process using this technique is 100% for relatively small batch samples. The superconducting properties, $T_c$, $J_c$ and trapped fields, of the single grains fabricated using the buffers are reported and the microstructures in the vicinity of the buffer of single grains fabricated by the modified technique are analysed.

Key words: buffer, SmBCO-Ag single grains, reliable seeding, silver, superconducting properties

## 1. Introduction

Bulk, single grain superconductors in the (RE)BCO family (where RE = Sm, Gd and Y) have considerable potential for practical applications due to their ability to trap magnetic fields up to ten times higher than the fields available using conventional permanent magnets [1]. Unfortunately, however, the oxide nature of (RE)BCO single grains means they tend to form brittle compounds and exhibit mechanical properties that are characteristic of ceramic materials. This is particularly significant given that the tensile forces associated with a high, trapped magnetic field can be hundreds of tonnes, which makes single grain bulk superconductors extremely susceptible to mechanical failure by fracture initiating from defects, pores and cracks that pre-exist in as-processed single grains. As a result, Ag has been used for some time to improve the mechanical properties of GdBCO single grains by in-filling the pores in the single grain microstructure generated during peritectic solidification [2][3][4]. However, the presence of Ag in the precursor pellet reduces the melting temperature of the single crystal seed in the top seeded melt growth (TSMG) process, which complicates significantly the controlled nucleation and subsequent epitaxial growth of a single grain, which is essential for high field applications. SmBCO has the highest superconducting transition temperature, $T_c$, and exhibits the most pronounced "peak" effect at higher magnetic field of all materials in the family of (RE)BCO bulk superconductors [5][6] and, therefore, has the greatest potential for use in practical applications (compared to GdBCO and YBCO, in particular). The reliability of the seeding process reduces significantly when Ag is introduced to the SmBCO system, in particular, where the melting temperature of the bulk composition is very close to that of the generic, NdBCO(MgO) seed. Failure of the seed is a direct cause of failure of the single grain growth process. Therefore, the development of an effective and reliable seeding technique is one of the greatest challenges in fabricating large quantities of SmBCO-Ag single grains. As a result, it is extremely important to increase the success rate of the TSMG process to batch-produce samples for commercial applications of these technologically important materials, and particularly so given the increasing cost and limited availability of rare-earth elements. In this paper, we report a novel technique that involves introducing a buffer layer between the seed and the precursor pellet that forms a physical barrier to inhibit the diffusion of Ag from the green body to the seed during melt processing in orderto prevent the melting of the seed. The use of the buffer solves completely the issue of seed melting, and increases the success rate of the seeding process to 100% of the samples fabricated in small batches by this process to date. Batch processed, single grains fabricated by this technique are presented, their microstructures in the vicinity of the buffer are analysed and their superconducting properties, including $T_c$, $J_c$ and trapped fields, are measured in order to demonstrate the positive effects of the buffers. A number of reasons for the apparent success of the novel seeding process are proposed.

## 2. Experimental

*2.1 A longstanding problem with seeded growth of SmBCO-Ag*

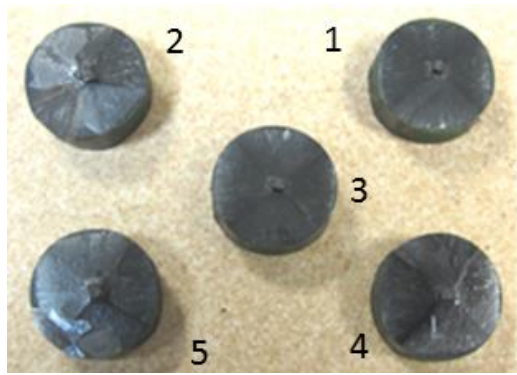

Figure 1. An example of seeding failure (sample 2)

Figure 1 shows five SmBCO-Ag samples that were grown together as a batch in a box furnace using a common TSMG process in an air atmosphere. Generic, NdBCO(MgO) seeds [7][8] were used to nucleate grain growth for each sample, as shown in fig. 1, from which it can be seen that the seed was used successfully to grow single grain SmBCO-Ag samples (labelled 1, 3, 4, and 5). Sample 2 (top left sample), however, failed to grow fully into a single grain due to a problem with the seed, and consists of multi-grains. The resulting seeding success rate of 80% achieved in this experiment, which is typical for a standard SmBCO batch process of this type, is not good enough for the large-scale fabrication of this material for commercial applications. SmBCO has a higher melting temperature than YBCO and GdBCO (the three most popular and potentially applicable (RE)BCO systems) and is therefore the closest to the melting temperature of the seed used in the melt-process. Adding Ag to SmBCO decreases the melting temperature of the seed by about 40 °C, which makes the melt processing of SmBCO-Ag single grains extremely challenging. The maximum temperature during processing, $T_m$, and holding time at $T_m$ were 1071 °C and 10 minutes, respectively. This value of $T_m$ corresponds roughly to the melting temperature of undoped SmBCO, which is necessarily low to avoid the failure of the seed ($T_m$ is usually required to be about 35 °C higher than the melting temperature of the SmBCO-Ag system to be seeded.).

*2.2 A new approach to seeded melt growth*

Higher $T_m$ and longer holding times are preferred in the TSMG of (RE)BCO to achieve complete peritectic decomposition of the superconducting $REBa_2Cu_3O_{7-\delta}$ (RE-123) phase to the $RE_2BaCuO_5$ (RE-211) and $Ba_3Cu_5O_8$ liquid phases (where RE = Sm, Gd and Y). However, the seed crystal is much more likely to fail when a higher $T_m$ and longer holding time is used for melt processing when Ag is present in the mixed precursor powder. A material buffer between the pellet and seed to isolate the seed from Ag in the precursor pellet during processing was therefore used in an attempt to maintain a relatively high decomposition temperature of the seed, at least during the nucleation and initial growth phase of the SmBCO/Ag single grain. Fig. 2 shows three arrangements of the SmBCO-Ag green pellet prior to melt processing, each using a generic MgO-NdBCO seed. In each case, the composition of the SmBCO-Ag pellet was (75wt% Sm-123 + 25wt% Sm-211) + 1wt% $CeO_2$ + 2wt % $BaO_2$ + 10wt % $AgO_2$. The commercial precursor powders had a purity of 99.9%. In total six pellets, each of mass 10 g, were prepared, again as shown in fig. 2, with three types of seed-precursor pellet arrangements and two buffer pellets investigated purposefully for this study. The first type of buffer was of composition 75wt% Sm-123 + 25wt% Sm-211, with a weight of 0.3 g and diameter of 5 mm. The second buffer consisted only of Sm-211, but, again, had a weight of 0.3 g and a diameter 5 mm. The third arrangement consisted only of generic seed placed in direct contact with the top surface of the pellet. The same heating profile was used for all six samples, with a $T_m$ and holding time at $T_m$ of 1071 °C and 20 minutes, respectively.

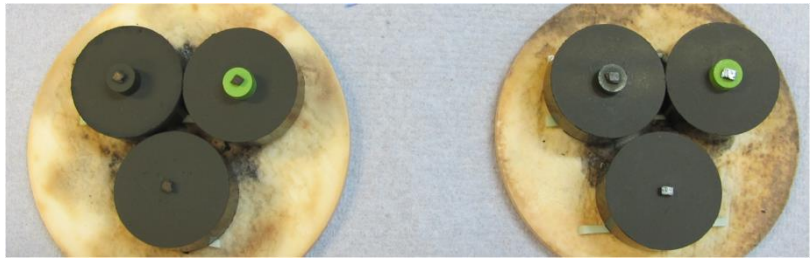

Figure 2. Two sets of three, different seed-precursor pellet arrangements. The Sm-211 (green phase) buffer is clear from the figure.

*2.3 Measurements of the superconducting properties, $T_c$, $J_c$ and trapped fields, of the SmBCO-Ag single grains batch-fabricated using buffers*

Three batches of SmBCO-Ag single grains of diameter 20 mm, 25 mm and 31mm were fabricated using the new buffering technique. $T_m$ and holding time for each heating process was maintained at 1071 °C and 20 minutes, respectively. The complete heating profile is shown in fig. 2, with the cooling rate over the temperature range between 1024 °C to 1004 °C reducing from 0.5 °C/hr to 0.2 °C/hr for increasing sample diameter from 20 mm to 31 mm. The fully grown single grains were annealed in 1% $O_2$ in Ar at 850 °C for 3 days [9] before being oxygenated in flowing $O_2$ (99.9%) at 380 °C for 10 days. The top surface of every sample was polished flat for trapped field measurements. Each sample was field-cooled (FC) to 77 K using liquid nitrogen in a field of 1.3 T applied perpendicular to its top surface. The applied field was then removed and the trapped field on the top surface of each sample measured using a rotating array of 20 Hall probes. The distance between the sample surface and the Hall probes was estimated to be 0.7 mm.

In order to estimate local $T_c$ and $J_c$, a randomly chosen SmBCO-Ag single grain of diameter 20 mm was cut into small specimens as illustrated schematically in fig. 3 (b) (taken from reference [10]). A MPMS XL SQUID magnetometer was used to measure $T_c$ and M–H loops for increasing and decreasing field cycles for each specimen. The extended Bean model [11] was used to estimate $J_c$ for the orthorhombic samples using the formula $J_c = 20m/(abc)/(a(1-a/3b))$ (Acm$^{-2}$), where $m$ (emu) represents the difference in magnetic moment observed in the M–H loops for increasing and decreasing field cycles with the applied field perpendicular to the ab-plane. Here a, b and c are the dimensions of the specimens along the principal crystallographic axes (where $a < b$), and are measured in centimetres.

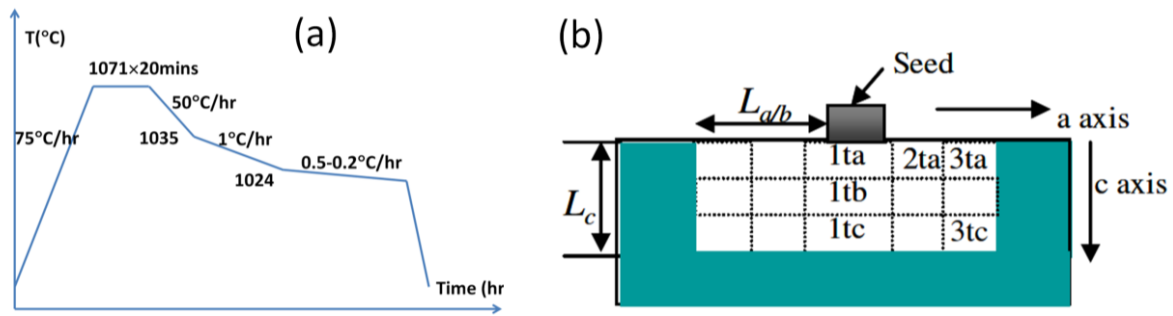

Figure 3. (a) Heating profiles used for batch-processing SmCO-Ag single grains of different sizes from 20 mm to 31 mm in diameter. (b) Schematic illustration of the location of the sub-specimens for characterisation of the superconducting properties in the parent single grain [10].

## 3. Results and discussion

*3.1 Reliability of buffering technique for fabricating SmBCO-Ag single grains*

Figure 4 shows that the results obtained from the TSMG processing of the two sets of three samples are the same. In each case, the buffer layers of composition of 75wt% Sm-123 + 25wt% Sm-211 survive the melt process for a $T_m$ of 1071 °C and a holding time of 20 minutes (the temperature was measured using a thermocouple located at the back of the furnace chamber, positioned approximately 5 cm away from each sample). The pellets seeded using the Sm-211 buffer and those seeded directly, however, all failed under the same processing conditions.

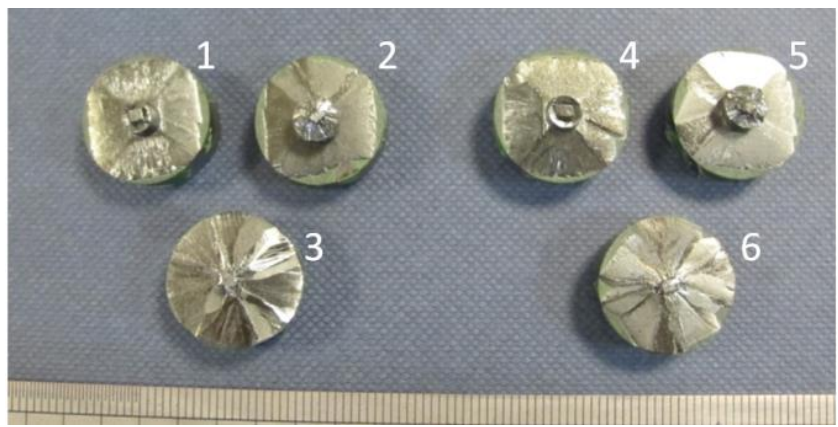

Figure 4. Photographs of the top surfaces of the samples grown using the different seed arrangements shown in fig. 2.

Figure 5 shows the microstructure of the sample cross-section along the crystallographic *c* direction (i.e. pellet thickness) and through the position of the seed of sample 1 (labelled in fig. 4). It can be seen that Ag (white in contrast in the photograph) is distributed throughout the sample, and mainly within the holes left by the residual porosity from the melt process. It can be seen further that Ag permeates into the buffer layer during melt processing, but terminates before it reaches the interface between the seed and the buffer. The inset photograph was taken at a magnification of 1000×, from which it can be seen more clearly that no Ag is present within the generic seed microstructure. As a result, the melting temperature of the seed and lattice structure were not altered during melt processing, enabling the generic seed to function well under these conditions of melt growth, even though the seed exhibits a relatively large amount of porosity.

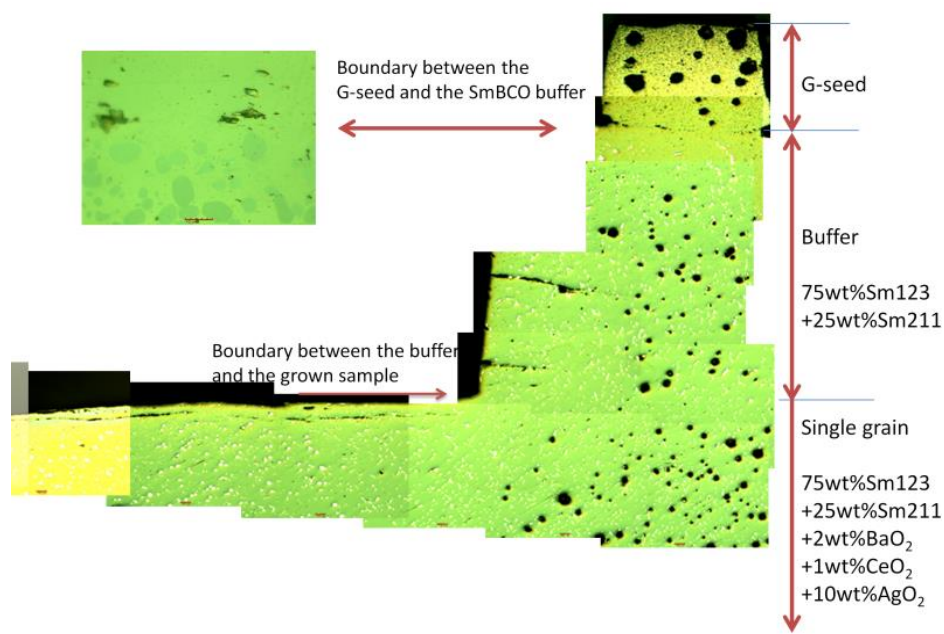

Figure 5. Microstructure of the cross section of sample 1 (shown in fig. 3) along the crystallographic *c* direction (i.e. pellet thickness) and through the position of the seed.

A buffer of composition of 75wt% Sm-123 + 25wt% Sm-211 is now used routinely by the Cambridge Bulk Superconductivity Group to process SmBCO-Ag using TSMG based on the results of this study, which has enabled large SmBCO-Ag (10wt% $Ag_2O$) single grains to be fabricated reliably. Figs 6(a) and (b) show samples of diameter 25 mm and 31 mm fabricated using the buffer layer technique. In total, about 50 single grain samples have been seeded recently using the buffer layer technique with a success rate of 100%. (Significantly, this technique can also be used to grow GdBCO-Ag single grains, which are equally easy to seed when a buffer layer is employed.)

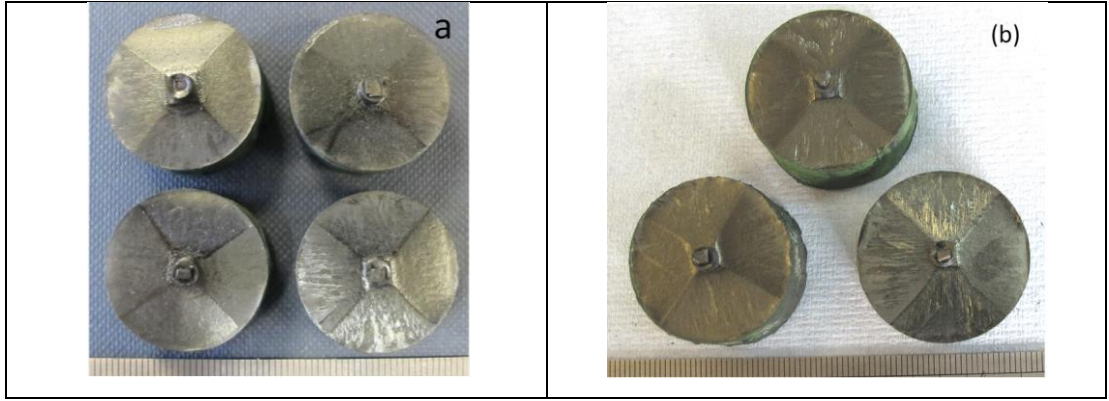

Figure 6. SmBCO-Ag single grain samples of diameter (a) 25 mm and (b) 31mm fabricated using a buffer layer.

*3.2 Superconducting properties SmBCO-Ag single grains fabricated using the buffering technique*

Figures 7 (a) and (b) show the measured $T_c$s and estimated values of $J_c$ using the extended Bean model. It can be seen that most of the $T_c$ curves exhibit a high onset $T_c$ (92-93 K) and a sharp superconducting transition, except for the data for specimens 2ta and 3ta. The $J_c$ curves for specimen positions other than 2ta and 3ta exhibit a characteristic shape, including a relatively high $J_c(0)$ and a peak effect at about 1.5 T. The low $J_c$ for specimens 2ta and 3ta is expected given the relatively low $T_c$ of these samples (the measurement temperature of 77 K is closer to the $T_c$ of the samples, which reduces $J_c$). The observed lower $T_c$ for specimens 2ta and 3ta suggests that Sm/Ba substitution is not suppressed sufficiently for these two positions in the parent single grain. Another motivation for using the buffers is to enhance the superconducting properties of the specimen under the seed (i.e. at position 1ta) where both $T_c$ and $J_c$ are typically low due to contamination from the seed and reduced flux pinning in this region [12][13]. It can be seen that the blue curve in fig. 7(a) has an onset $T_c$ of 92.5 K and exhibits a rather sharp transition, although it is not the best $T_c$ curve measured. The $J_c$ curve of specimen 1ta (light purple) is also reasonable with a $J_c(0)$ of $5.0 \times 10^4$ A/cm$^2$ and a peak effect at 1 T. The observed values of $T_c$ and $J_c$ for the specimen located immediately below the seed indicate that buffers used in this research not only prevent Ag infiltrating the seed to enhance the reliability of the seeding process, but also limit the harmful effects of the elements Nd and Mg diffusing from the seed to the single grain. A direct consequence of this is that the superconducting properties of the SmBCO-Ag single grain specimen under the position of the seed are increased to the level reported in previous studies [12][13]. The differences in $T_c$ and $J_c$ of the nine specimens suggest that Sm/Ba substitution is not totally supressed in the research. However, the buffer technique clearly has no negative effects on the superconducting properties of the seeded samples.

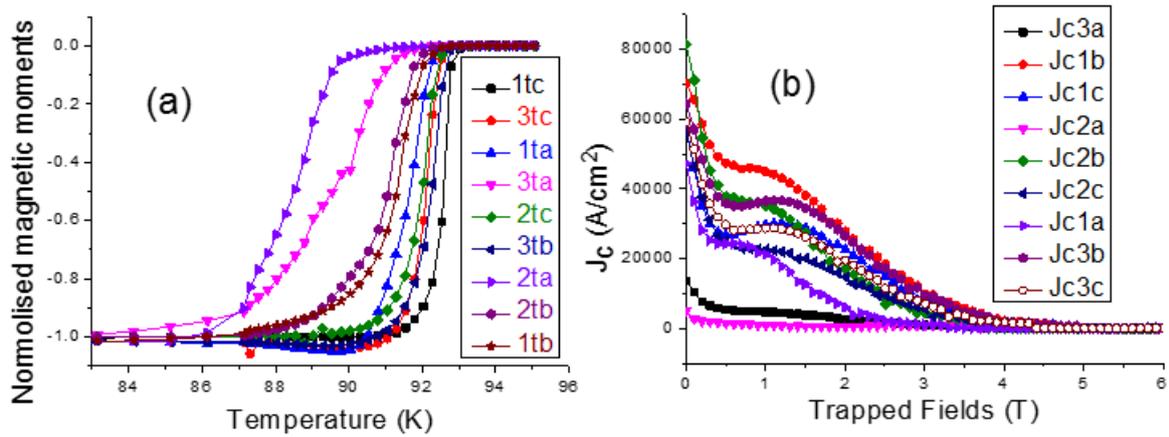

Figure 7. The superconducting properties of nine specimens cut from different positions in a SmBCO-Ag single grain fabricated using a buffer layer (a) $T_c$ measurements (b) $J_c$ estimations

Figure 8 shows photographs of the top surfaces of five polished SmBCO-Ag single grains. The trapped fields of these samples measured at 77 K are indicated in the figure. These values are higher than the average value (0.5 T) observed typically for YBCO single gains of these dimensions, but lower than the average value (0.7 T) observed for GdBCO-Ag single grains. Although the values of trapped field obtained in this research are the highest in the SmBCO-Ag system for samples this diameter, there is still potential to improve its superconducting properties by further suppression of Sm/Ba substitution effects.

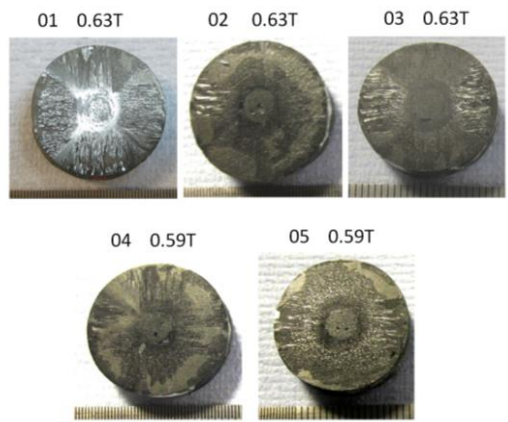

Figure 8. Photographs of the top surface of five SmBCO-Ag single grains of diameter 20 mm. The trapped fields are as indicated.

Figure 9 shows the average trapped field profiles of the SmBCO-Ag single grains of diameter 25 mm and 31 mm, shown in figs 6(a) and (b). Average values 0.60 T for the 25 mm diameter samples and 0.65 T for 31 mm diameter samples are obtained. The variations in the average value of trapped field across all the samples of a given diameter are shown in the figure. Although a single grain of diameter 31 mm is the largest SmBCO-Ag single grain reported to date, the observed magnitude of trapped field for these lager samples is significantly lower than is expected for a fully superconducting sample of this size. The reason for the observed low values of trapped field is the severe effect on $T_c$ of Sm/Ba substitution that occurs when melt processing in performed in an air atmosphere. However, the necessary annealing process in 1%$O_2$ in Ar and subsequent oxygenation have not yet been optimised for large samples, and it is anticipated that increased trapped fields will be produced when this has been performed.

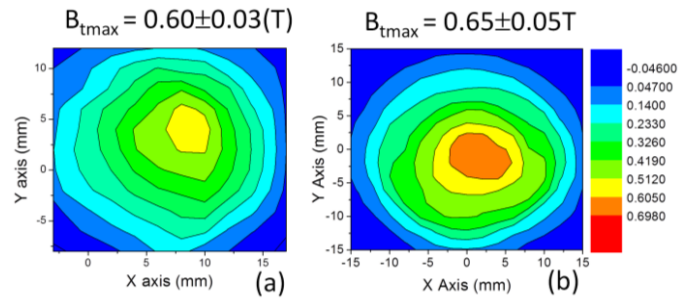

Figure 9. Average trapped fields of the SmBCO-Ag single grains of diameter (a) 25 mm and (b) 31 mm.

*3.3 Explanation of the function of the buffer*

Several attempts to use a buffer layer have been made previously in the TSMG process [12][13] with the specific aim of preventing contamination of the precursor pellet with Nd or Sm from the seed positioned on the upper surface of the buffer. However, the buffer layer technique developed here focuses on inhibiting the diffusion of Ag (Ag is more mobile and therefore more difficult in control) into the single crystal seed from the bottom of the sample during melt processing, which is an extension of these earlier studies. The buffers used in this study also function to inhibit diffusion of impurities (Nd, Mg and Ag) from both the top and bottom of the buffer, so that the seed can survive under relatively severe processing conditions to enable reliable batch processing of SmBCO-Ag.

Zhou et al suggested that buffers can reduce the lattice mismatch between the seed and the precursor pellet [14, 15], and this is supported by the results of the current study. It is apparent that the lattice mismatch between the seed and the grown single grain can be very small if only the parameters of the respective single cells are compared (in the present case, the difference in lattice parameters *a, b,* and *c* between the generic seed and grown single grain SmBCO-Ag is of the order of 0.1 Å). However, the difference in length caused by lattice mismatch in the seeding region (typically ~2×2 mm$^2$) can increase to several tenths of one mm in one dimension, which can cause failure of the seeding process. When a buffer is used, this accumulated lattice mismatch can be overcome by expanding the buffer layer volume into the processing atmosphere (i.e. obstacle free and air in this investigation). In this case, the grown buffer functions subsequently as a more effective seed in terms of impurity content and lattice compatibility, given that its composition is very close to that of the target single grain. Fig. 10 shows two, randomly chosen buffer layers and their seeds used to melt-process the single grain samples described in this study, from which it can be seen clearly that each buffer has, itself, grown into a single grain. Facet lines are clearly visible on both the top and side surfaces of these samples, which is indicative generally of successful top seeded melt growth. These fully-grown single grains with buffers are structurally better than those grown directly with a generic seed in terms of lattice compatibility with the SmBCO-Ag pellet. As a result, SmBCO-Ag single grains can been grown easily and reliably using a buffer layer.

Experience based on the growth of over a hundred GdBCO-Ag and SmBCO-Ag single grain samples, with and without buffer layers, suggests that the buffer layer technique also aids the seeding process itself. The buffer technique, which is well known in the field of crystal growth [16][17], effectively stabilises the seed and filters impurities from the seeding and growth process [16][17]. The buffers used in this study functioned additionally as seeds with fewer defects, which suggests the buffering technique is potentially applicable to other (RE)BCO systems for increasing reliability of the TSMG process. It may be concluded, therefore, that a buffer layer of composition 75wt% Sm-123 + 25wt% Sm-211 has been identified to stabilise the decomposition temperature of the generic Mg-NdBCO seed and acts as an effective barrier to impurities during processing, such as Nd, Mg, Ag and $CeO_2$ in the present case, which are potentially harmful to the seeding process. Further research, however, is needed to investigate and understand the mechanism in more detail.

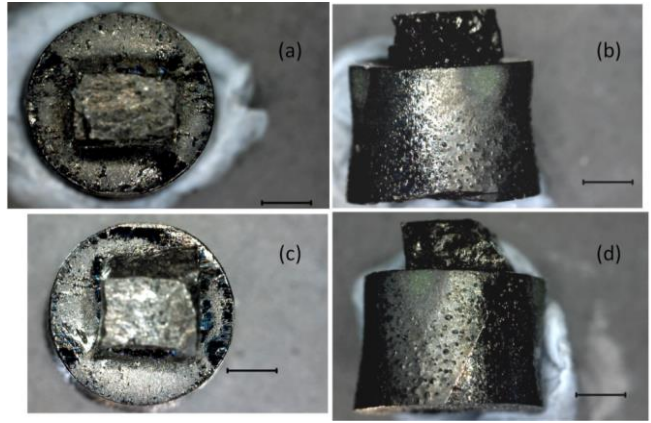

Figure 10. Photographs of buffer layers and their seeds after melt processing (the scale bar represents 1 mm).

## 4. Conclusions

A buffer layer of composition of 75wt%Sm-123 + 25wt%Sm-211, diameter 5 mm and weight 0.3 g has been identified as an effective aid to the seeding process using a generic NdBCO(MgO) single crystal seed. A success rate of 100% of the seeding process has been achieved using the buffer layer technique, which is an essential requirement for fully growing single grains on a large scale. Large, SmBCO-Ag single grains of diameter up to 31 mm have been batch-fabricated successfully using this technique. Analysis of the sample microstructure and the features of the buffer layer and seed after melt processing suggests that the presence of the buffer layer stabilises the seed and presents an effective physical barrier to contamination of the single crystal seed by impurity elements, and by Ag, in particular, in the present study. The superconducting properties of the SmBCO-Ag single grains are not affected by the buffer-seeding process. Record trapped fields have obtained in the SmBCO-Ag system using the buffer layer, although further research is required to suppress Sm/Ba substitution in the TSMG process for these materials fabricated in an air atmosphere.

## References

[1] Durrell J H, Dennis A R, J Jaroszynski, Ainslie M D, Palmer K G B, Shi Y-H, Campbell A M, Hull J, Strasik M, Hellstrom E E and Cardwell D A 2014 S*upercond. Sci. Technol.* **27** 082001
[2] Nariki S, Sakai N and Murakami M 2005 *Supercond. Sci. Technol.* **18** S126
[3] Shi Y, N Hari Babu, Iida K, Yeoh W.K, Dennis A R, Pathak S K and Cardwell D A 2010 *Phys. C: Supercond.* **470** 685
[4] Muralidhar M, Tomita M, Suzuki K, Jirsa M, Fukumoto Y and Ishihara A 2010 *Supercond. Sci. Technol.* **23** 045033
[5] Murakami M, Yoo S, Higuchi T and Sakai N 1994 *Jpn. J. Appl. Phys.*, **33** L715.
[6] N. Hari Babua, Iida K, Shi Y and Cardwell D A 2008 *Physica C: Superconductivity,* **468** 1340
[7] Shi Y, N Hari Babu and D A Cardwell 2005 *Supercond. Sci. Technol.* **18** L13–L16

[8] Nadendla H B, Shi Y, Iida K and Cardwell D A 2005 *Nature Materials* **4** 476-480
[9] Shi Y, Desmedt M, Durrell J, Dennis A R and Cardwell D A 2013 *Supercond. Sci. Technol.* **26** 095012
[10] Shi Y, N. Hari Babu, Iida K and Cardwell D A 2007 *Supercond. Sci. Technol.* **20** 38
[11] Chen D-X and Goldfarb R B 1989 *J. Appl. Phys.* **66** 2489
[12] Kim C-J, Lee J H, Park S-D, Jun B-H, Han S C and Han Y H 2011 *Supercond. Sci. Technol.* **24** 015008
[13] Li T Y, Cheng L, Yan S B, Sun L J, Yao X, Yoshida Y and Ikuta H 2010 *Supercond. Sci. Technol.* **23** 125002
[14] Zhou D-F, Xu K, Hara S, Li B-Z and Izumi M 2013 *Trans. Nonferrous Met. Soc. China* **23** 2042
[15] Zhou D-F, Xu K, Hara S, Li B-Z, Deng Z, Tsuzuki K and Izumi M 2012 *Supercond. Sci. Technol.* **25** 025022 *Biology* **142** 66
[17] Stewart P D S, Kolek S A, Briggs R A, Chayen N E and Baldock P F M. Random 2011 *Cryst. Growth Des.* **11 (8)** 3432